\documentclass[conference]{IEEEtran}
\IEEEoverridecommandlockouts

\usepackage{cite}
\usepackage{amsmath,amssymb,amsfonts}
\usepackage{algorithmic}
\usepackage{graphicx}
\usepackage{textcomp}
\usepackage{xcolor}
\usepackage{forest}
\usetikzlibrary{trees,positioning,shapes,shadows,arrows.meta}
\usepackage{tikz}

\makeatletter
\newcommand{\linebreakand}{%
  \end{@IEEEauthorhalign}
  \hfill\mbox{}\par
  \mbox{}\hfill\begin{@IEEEauthorhalign}
}
\makeatother

\begin{document}

\title{From Text to Multimodality: Exploring the Evolution and Impact of Large Language Models in Medical Practice}

\author{
    \IEEEauthorblockN{
        Qian Niu\textsuperscript{1}, 
        Keyu Chen\textsuperscript{2},
        Ming Li\textsuperscript{2},
        Pohsun Feng\textsuperscript{3},
        Ziqian Bi\textsuperscript{4},
        Lawrence KQ Yan\textsuperscript{5},
        Yichao Zhang\textsuperscript{6},\\
        Caitlyn Heqi Yin\textsuperscript{7},
        Cheng Fei\textsuperscript{8},
        Junyu Liu\textsuperscript{1},
        Tianyang Wang\textsuperscript{9},
        Yunze Wang\textsuperscript{10},
        Silin Chen\textsuperscript{11},\\
        Ming Liu\textsuperscript{12},
        Benji Peng\textsuperscript{2}
        Xinyuan Song\textsuperscript{13},
        Ziyuan Qin\textsuperscript{13},
        Riyang Bao\textsuperscript{13},
        Zekun Jiang\textsuperscript{*,14}
    }
    \IEEEauthorblockA{
        \textsuperscript{1}Kyoto University
    }
    \IEEEauthorblockA{
        \textsuperscript{2}Georgia Institute of Technology
    }
    \IEEEauthorblockA{
        \textsuperscript{3}National Taiwan Normal University
    }
    \IEEEauthorblockA{
        \textsuperscript{4}Indiana University
    }
    \IEEEauthorblockA{
        \textsuperscript{5}Hong Kong University of Science and Technology
    }
    \IEEEauthorblockA{
        \textsuperscript{6}The University of Texas at Dallas
    }
    \IEEEauthorblockA{
        \textsuperscript{7}University of Wisconsin-Madison
    }
    \IEEEauthorblockA{
        \textsuperscript{8}Cornell University
    }
    \IEEEauthorblockA{
        \textsuperscript{9}University of Liverpool
    }
    \IEEEauthorblockA{
        \textsuperscript{10}University of Edinburgh
    }
    \IEEEauthorblockA{
        \textsuperscript{11}Zhejiang University
    }
    \IEEEauthorblockA{
        \textsuperscript{12}Purdue University
    }
    \IEEEauthorblockA{\textsuperscript{13}Emory University, Atlanta, GA, USA}
    \IEEEauthorblockA{\textsuperscript{14}West China Biomedical Big Data Center, West China Hospital, Sichuan University, Chengdu, China}
    \IEEEauthorblockA{
        *Corresponding Email: zekun\_jiang@163.com
    }
}

\maketitle


\begin{abstract}
Large Language Models (LLMs) have rapidly evolved from text-based systems to multimodal platforms, significantly impacting various sectors including healthcare. This comprehensive review explores the progression of LLMs to Multimodal Large Language Models (MLLMs) and their growing influence in medical practice. We examine the current landscape of MLLMs in healthcare, analyzing their applications across clinical decision support, medical imaging, patient engagement, and research. The review highlights the unique capabilities of MLLMs in integrating diverse data types, such as text, images, and audio, to provide more comprehensive insights into patient health. We also address the challenges facing MLLM implementation, including data limitations, technical hurdles, and ethical considerations. By identifying key research gaps, this paper aims to guide future investigations in areas such as dataset development, modality alignment methods, and the establishment of ethical guidelines. As MLLMs continue to shape the future of healthcare, understanding their potential and limitations is crucial for their responsible and effective integration into medical practice.
\end{abstract}


\section{Introduction}
The landscape of healthcare is constantly evolving, driven by an unprecedented explosion of data. Electronic health records, medical imaging, genomic sequencing, and wearable sensors generate an overwhelming amount of information, exceeding human capacity for efficient analysis and interpretation \cite{Jiang2023Health}. This phenomenon presents both an opportunity and a challenge: ingesting this information can revolutionize healthcare, but doing so requires innovative tools capable of processing and synthesizing these diverse data streams. Artificial intelligence (AI) has emerged as a powerful force in addressing this challenge, with large language models (LLMs) at the forefront of this revolution.

Initially, LLMs focused primarily on text-based tasks, demonstrating remarkable proficiency in understanding and generating human-like language \cite{Singhal2023Large}. However, the inherent multimodality of medicine, where clinical decisions often rely on the synthesis of information from diverse sources such as images, text, and genomics, necessitates more versatile models \cite{Acosta2022Multimodal}. This need has given rise to Multimodal Large Language Models (MLLMs), a new generation of LLMs capable of processing and integrating information from various modalities. These advanced models potentially unlock a new era of precision medicine and personalized healthcare, offering a more comprehensive approach to medical data analysis and decision-making.

A key strength of MLLMs is their ability to bridge the gap between unstructured and structured data, a particularly valuable feature in healthcare where information is often fragmented across different formats. For example, the REALM framework leverages LLMs to encode clinical notes and integrates them with time-series EHR data, enhancing clinical predictions by incorporating external knowledge from knowledge graphs \cite{Yinghao2024REALM}. In a similar vein, the MedDr model \cite{He2024MedDr} employs a diagnosis-guided bootstrapping strategy to build vision-language datasets, showcasing superior performance across various medical tasks through a retrieval-augmented diagnosis approach. These advancements underscore the potential of MLLMs to enhance data interoperability and extract meaningful insights from diverse sources, potentially revolutionizing how healthcare professionals access and utilize patient information.

MLLMs show great potential for transforming healthcare by enabling a more comprehensive understanding of patient health, potentially leading to improved diagnostics, personalized treatment plans, and enhanced patient engagement \cite{chaves2024training}. For instance, these models could analyze a patient's medical history, imaging scans, and genetic data to provide more accurate diagnoses and predict disease risks, facilitating early intervention and tailored treatment strategies. In the field of medical imaging, the integration of LLMs has demonstrated significant progress. Research has shown the effectiveness of visual language models (VLMs), a subset of MLLMs, in analyzing various biomedical images, including brain MRIs, blood cell images, and chest X-rays \cite{Van2024On}. A notable example is the LlaVA-Rad model, a lightweight and open-source multimodal system that has achieved state-of-the-art results on standard radiology tasks. This model has surpassed larger counterparts in both performance and accessibility, making it particularly suitable for real-world clinical applications \cite{chaves2024training}.

MLLMs could also enhance communication between patients and healthcare providers through interactive chatbots and virtual assistants, potentially improving patient engagement and healthcare accessibility \cite{Mesko2023Impact}. The creation of chatbots like MedAide, which utilize optimized tiny-LLMs on edge devices, demonstrates the capacity of MLLMs to provide medical assistance in resource-limited settings and remote areas, addressing challenges in healthcare access \cite{Abdul2024MedAide}. However, developing reliable and trustworthy medical chatbots requires addressing critical issues such as accuracy, privacy protection, and bias mitigation to meet the high standards required for patient care and safety.

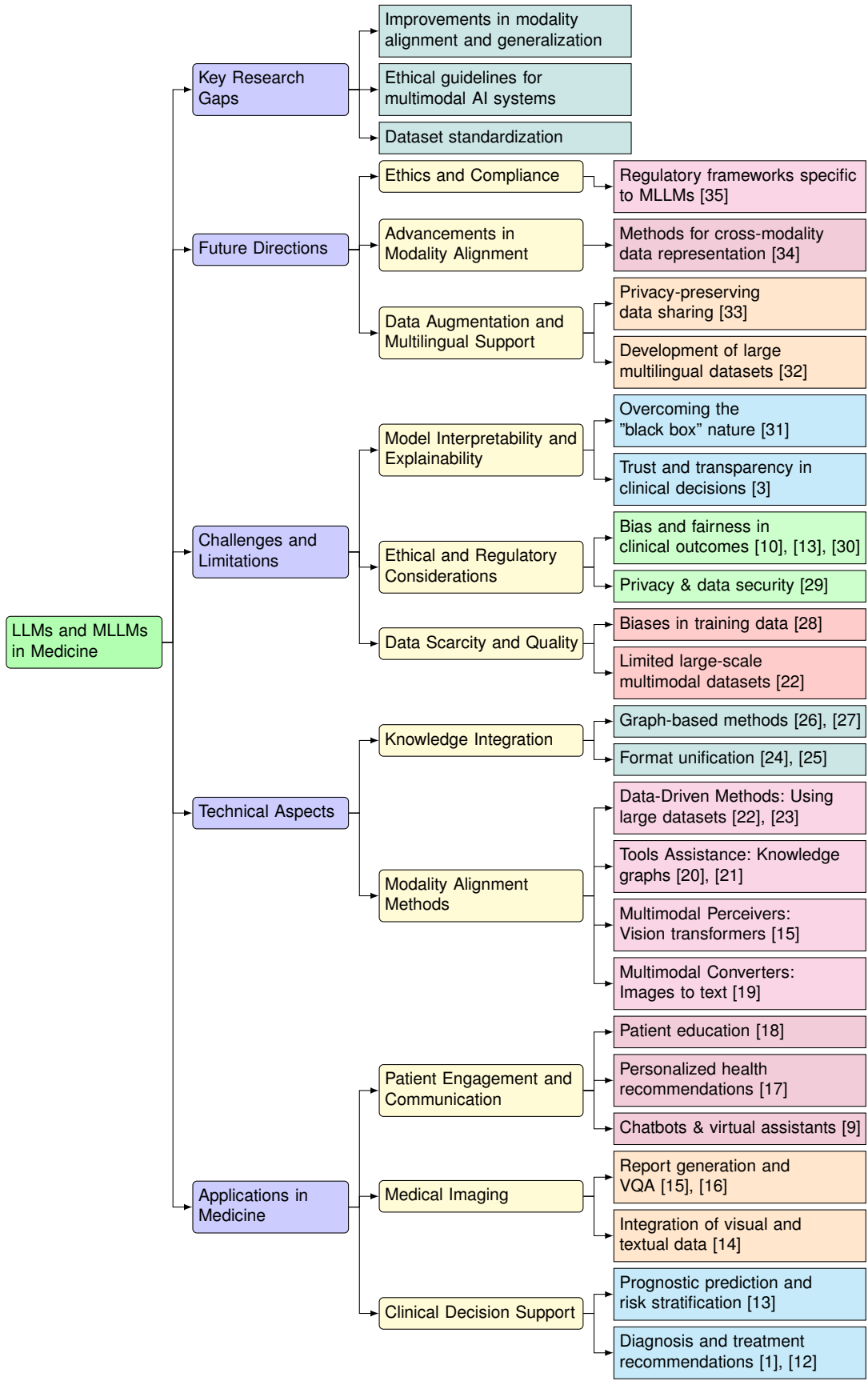
\begin{figure*}
    \centering
    
\tikzset{
    basic/.style  = {draw, align=left, font=\sffamily, rectangle},
    root/.style   = {basic, rounded corners=2pt, thin, fill=green!30, text width=3cm, anchor=west},
    level1/.style = {basic, thin, rounded corners=2pt, fill=blue!20, text width=3cm},
    level2/.style = {basic, thin, rounded corners=2pt, fill=yellow!20, text width=4cm},
    leaf1/.style   = {basic, thin, fill=red!20, text width=5cm},
    leaf2/.style   = {basic, thin, fill=green!20, text width=5cm},
    leaf3/.style   = {basic, thin, fill=cyan!20, text width=5cm},
    leaf4/.style   = {basic, thin, fill=orange!20, text width=5cm},
    leaf5/.style   = {basic, thin, fill=purple!20, text width=5cm},
    leaf6/.style   = {basic, thin, fill=magenta!20, text width=5cm},
    leaf7/.style   = {basic, thin, fill=teal!20, text width=5cm},
    edge from parent/.style={draw=black, edge from parent fork right},
    level distance=1.5cm,
}

\begin{forest}
for tree={
    grow=east,
    scale=0.8, 
    growth parent anchor=west,
    parent anchor=east,
    child anchor=west,
    l sep=5mm, 
    s sep=1mm, 
    edge path={\noexpand\path[\forestoption{edge},->, >={latex}] 
         (!u.parent anchor) -- +(5pt,0pt) |-  (.child anchor) 
         \forestoption{edge label};},
    align=left, 
}
[LLMs and MLLMs\\ in Medicine, root,
  [Applications in\\Medicine, level1,
    [Clinical Decision Support, level2,
      [Diagnosis and treatment \\ recommendations \cite{Jiang2023Health, Hiesinger2023Almanac}, leaf3]
      [Prognostic prediction and\\ risk stratification \cite{Zekai2023Language}, leaf3]
    ]
    [Medical Imaging, level2,
      [Integration of visual and\\ textual data \cite{Jiawei2024MISS}, leaf4]
      [Report generation and\\ VQA \cite{Michael2023Med, Zhou2024Pre}, leaf4]
    ]
    [Patient Engagement and\\ Communication, level2,
      [Chatbots \& virtual assistants \cite{Abdul2024MedAide}, leaf5]
      [Personalized health\\ recommendations \cite{dao2024llm}, leaf5]
      [Patient education \cite{Akash2024MedSumm}, leaf5]
    ]
  ]
  [Technical Aspects, level1,
    [Modality Alignment\\ Methods, level2,
      [Multimodal Converters:\\ Images to text \cite{Feilong2023X}, leaf6]
      [Multimodal Perceivers:\\ Vision transformers \cite{Michael2023Med}, leaf6]
      [Tools Assistance: Knowledge\\ graphs \cite{gao2023leveraging, Remy2024BioLORD}, leaf6]
      [Data-Driven Methods: Using\\ large datasets \cite{Emre2024Hippocrates, Kai2023BiomedGPT}, leaf6]
    ]
    [Knowledge Integration, level2,
      [Format unification \cite{shao2024dual, Qinghao2023mPLUG}, leaf7]
      [Graph-based methods \cite{niu2024ehr, ektefaie2023multimodal}, leaf7]
    ]
  ]
  [Challenges and\\ Limitations, level1,
    [Data Scarcity and Quality, level2,
      [Limited large-scale\\ multimodal datasets \cite{Emre2024Hippocrates}, leaf1]
      [Biases in training data \cite{Nassiri2024Recent}, leaf1]
    ]
    [Ethical and Regulatory\\ Considerations, level2,
      [Privacy \& data security \cite{peng2024securinglargelanguagemodels}, leaf2]
      [Bias and fairness in\\ clinical outcomes \cite{Harrer2023Attention, Zekai2023Language, Jiang2024754}, leaf2]
    ]
    [Model Interpretability and\\ Explainability, level2,
      [Trust and transparency in\\ clinical decisions \cite{Acosta2022Multimodal}, leaf3]
      [Overcoming the\\ "black box" nature \cite{Williams2024Use}, leaf3]
    ]
  ]
  [Future Directions, level1,
    [Data Augmentation and\\ Multilingual Support, level2,
      [Development of large \\ multilingual datasets \cite{Iker2024MedMT5}, leaf4]
      [Privacy-preserving\\ data sharing \cite{Wang2023Accelerating}, leaf4]
    ]
    [Advancements in\\ Modality Alignment, level2,
      [Methods for cross-modality\\ data representation \cite{Zhou2024Generalist}, leaf5]
    ]
    [Ethics and Compliance, level2,
      [Regulatory frameworks specific\\ to MLLMs \cite{Ong2024Ethical}, leaf6]
    ]
  ]
  [Key Research\\ Gaps, level1,
    [Dataset standardization, leaf7]
    [Ethical guidelines for\\ multimodal AI systems, leaf7]
    [Improvements in modality\\ alignment and generalization, leaf7]
  ]
]
\end{forest}
    \caption{Simplified Taxonomy of MLLMs in Medicine}
    \label{fig:taxonomy}
\end{figure*}

Our review aims to offer an overview of the current state of MLLMs in medicine and healthcare. We will not only examine their architecture, capabilities, and limitations, but also explore potential applications across various medical domains. We will critically assess the challenges and research gaps impeding the widespread adoption of MLLMs in clinical settings, including data limitations, technical difficulties, and ethical considerations \cite{Harrer2023Attention}. For example, the evaluation of LLMs in healthcare often relies on benchmarks that are usually unfit for real-world diagnostic frameworks and are likely vulnerable to data leakage \cite{Lyu2024Language}, which indicates the need for standardized evaluation frameworks and comprehensive datasets that accurately reflect the clinical practice. By analyzing the current research landscape and identifying key areas for further development, this review seeks to guide the responsible and effective integration of MLLMs into healthcare. Our goal is to contribute to a brighter future where AI assists clinicians and enhances patient care, while addressing the unique challenges and requirements within the field. In order to provide a clear overview of the various applications and components of MLLMs in medicine, we present a taxonomy \textbf{Fig \ref{fig:taxonomy}}. This simplified taxonomy categorizes the key aspects of MLLMs in healthcare and medicine.

\section{The Rise of Multimodal Large Language Models in Medicine}
\subsection{LLMs and Their Evolution: From Text to Multimodal Understanding}

Large language models (LLMs) represent a significant advancement in artificial intelligence, demonstrating remarkable capabilities in comprehending and generating human-like text. Architecturally, they often rely on the Transformer network \cite{vaswani2017attention}, a powerful neural network structure that excels at capturing long-range dependencies and contextual relationships within text. LLMs are initially trained on massive text corpora, a process known as pre-training, to develop a generalized understanding of language structure and patterns. This pre-training phase allows them to learn a wide range of linguistic features and relationships, making them adaptable to various downstream tasks.

LLMs can be fine-tuned on smaller, task-specific datasets to further refine their performance in specific domains. For example, ClinicalT5 \cite{Lu2022ClinicalT5} demonstrates how a general-purpose LLM (T5) can be adapted for clinical text by fine-tuning it on the MIMIC-III dataset. This adaptation to the medical domain is crucial for addressing the unique challenges of medical language, including its specialized vocabulary and complex semantic relationships \cite{Meng2024application}.

Despite impressive capabilities, LLMs may face many limitations. One notable issue is "hallucination", where the model generates plausible but incorrect or nonsensical information, as highlighted in the study by Ziaei and Schmidgall \cite{Rojin2023Language}. Hallucination can be particularly problematic in healthcare, where accuracy and reliability are the top priorities \cite{Nassiri2024Recent, peng2024emerging}. Additionally, biases present in the training data can propagate to the model's outputs, leading to unfair or discriminatory outcomes, as discussed in the paper by Reddy \cite{Reddy2023Evaluating}. Addressing these biases requires careful data curation and model development strategies \cite{han2024medsafetybenchevaluatingimprovingmedical}.

Integrating LLMs with other modalities, such as images and videos, results in MLLMs. MLLMs like GPT-4V \cite{Deng2024When, achiam2023gpt} and Gemini \cite{Ankit2024Gemini, team2023gemini} process and generate both text and visual information, which opens up new possibilities for medical applications. For example, MLLMs can be used to generate captions for medical images \cite{Iryna2024Vision}, answer visual questions about medical images \cite{Yutao2024OmniMedVQA}, and even assist in medical report generation \cite{Nakaura2024impact}. On the other hand, MLLMs are still in their early stages, and these models often face challenges in terms of accuracy, reliability, and ethical considerations \cite{Yihe2024Unbridled}. Further research is needed to fully explore the potential of MLLMs and address these challenges to enable their safe and effective deployment in clinical practice.

\subsection{Multimodality in Medicine: Embracing the Rich Variety of Data}

Medicine is inherently multimodal, as it involves many types of information beyond just written text. For example, when a patient comes in with a possible lung infection, their case might include several kinds of data: written information like their medical history and symptoms noted by doctors, images from chest X-rays, sound recordings of their breathing, and even genetic information to assess their personal risk. Combining these different types of information is important for getting a complete picture of a patient's health and more accurate and personalized medical care \cite{Tripathi2024Efficient}. This is where multimodal models shine, because they are designed to process and integrate various types of data.

We have seen a surge in developing MLLMs capable of processing and integrating diverse medical data types \cite{Duzhen2024MM}. The study by Tian et al. \cite{Liu2023medical} exemplifies this trend, showcasing a Med-MLLM model that handles both visual and textual data for improved clinical decision-making, particularly in rare diseases and pandemics. MLLMs could revolutionize various medical practices. For instance, in radiology, MLLMs are being explored for generating comprehensive reports \cite{Bhayana2024Chatbots}, assisting in diagnosis by analyzing both images and clinical notes \cite{Yinghao2024REALM}, and facilitating visual search and querying within patient imaging history \cite{Yildirim2024Multimodal}.

MLLMs for more specialized medical tasks has also gained momentum. SkinGPT-4 is a system designed for dermatological diagnosis using both images and clinical data, which offers autonomous image evaluation and treatment recommendations \cite{Zhou2024Pre}. Developing robust and reliable MLLMs requires overcoming many challenges. Large, diverse, and unbiased medical datasets across multiple modalities are crucial \cite{Yutao2024OmniMedVQA}. Accuracy, interpretability, explainability, interoperability, and ethics are important to be discussed before integrating into existing clinical workflows \cite{Ong2024Ethical}.

\subsection{Modality Alignment Methods: Bridging the Semantic Gap}

Integrating different data types into LLMs is challenging, mainly because of differences in how each type represents information. Aligning these modalities is essential for LLMs to process and reason over multimodal data. Researchers are currently exploring several methods for addressing this issue, which can be grouped into four main categories.

\begin{itemize}
    \item \textbf{Multimodal Converters}: These methods transform data from different modalities into a unified representation that LLMs can understand. For example, images might be converted into textual descriptions or embeddings before being fed into the LLM. This approach is seen in models like X-LLM \cite{Feilong2023X}, which treats modalities as foreign languages and converts them to text, or LIFTED \cite{Wenhao2024Multimodal}, which transforms modalities into natural language descriptions for improved clinical trial outcome prediction.
    \item \textbf{Multimodal Perceivers}: These methods directly enhance the LLMs' perception of multimodal data. A vision encoder can be integrated into the LLM architecture to enable it to directly process and understand images and texts. Med-Flamingo \cite{Michael2023Med} incorporates a vision transformer for medical image understanding. Similar approaches can be seen in models like SkinGPT-4 \cite{Zhou2024Pre}, which combines a vision transformer with a LLM for dermatological diagnosis, and MedVersa \cite{Zhou2024Generalist}, which uses a LLM as a learnable orchestrator to process both visual and linguistic information to interpret medical images.
    \item \textbf{Tools Assistance}: These methods uses external tools for multimodalities. A knowledge graph can link concepts across modalities and provide additional context for the LLM. The study by Gao et al. \cite{gao2023leveraging} uses the Unified Medical Language System (UMLS) knowledge to enhance diagnosis generation. A similar approach is used in BioLORD-2023 \cite{Remy2024BioLORD}, which integrates LLMs with knowledge graphs to improve performance in semantic textual similarity, biomedical concept representation, and named entity linking.
    \item \textbf{Data-Driven Methods}: These methods rely on large-scale multimodal datasets to train LLMs directly on multimodal tasks and often involves developing new architectures and training strategies so the modals can learn from different modalities simultaneously. Models like BiomedGPT \cite{Kai2023BiomedGPT} are trained on those diverse multimodal datasets. The recent open-source frameworks like Hippocrates \cite{Emre2024Hippocrates} further facilitates this approach by providing access to training datasets, codebases, and evaluation protocols, encouraging further collaborative efforts.
\end{itemize}

Each method has its own strengths and weaknesses. Multimodal converters are relatively simple but may lead to information loss during conversion \cite{park2023cross}. Multimodal perceivers can potentially capture richer relationships between modalities, but requires more complex architectures and training. Tools assistance uses existing knowledge bases and resources but may not be comprehensive or up-to-date. Data-driven methods can achieve high performance but require large and diverse datasets.

\section{Applications of MLLMs in Medicine}

\subsection{A. Clinical Decision Support}
While MLLMs integrate diverse data modalities, offering a more comprehensive view of patient health and the ability to detect complex patterns for improved diagnosis, treatment personalization, and risk assessment \cite{Tripathi2024Large}, their development is still in the early stages. As a result, LLMs continue to dominate the field due to their maturity and established performance. This section introduces both LLMs and MLLMs, while emphasizing the promise of multimodal models.

\textbf{Diagnosis and Treatment Recommendations}: NYUTron is an LLM trained on clinical notes, for predicting patient outcomes with high accuracy \cite{Jiang2023Health}. PMC-LLaMA is a performant LLM for medical Q\&A  \cite{Chaoyi2023PMC, Chaoyi2023PMC1, Wu2024PMC}. Almanac is an LLM augmented with retrieval capabilities from curated medical resources and has significant improvements in factuality, completeness, user preference, and safety for clinical decision-making \cite{Hiesinger2023Almanac}. Med-PaLM 2 is a specialized LLM for medicine, showcased superior performance on medical question answering and treatment recommendation tasks, significantly outperforming GPT-3.5 \cite{Singhal2023Large}. Med-PaLM M is a MLLM achieving competitive performance on medical question answering, radiology report generation, etc. \cite{Tu2023Towards}. 

\textbf{Prognostic Prediction and Risk Stratification}: Beyond diagnosis, LLMs have also shown promise in prognostic prediction and risk stratification. Researchers have tried to use LLMs for prognostic prediction in immunotherapy, achieving encouraging results in improving accuracy and facilitating early disease detection \cite{Zekai2023Language}. Studies have demonstrated the potential of LLMs and MLLMs to predict outcomes like mortality, length of stay, and readmission using structured EHR data, outperforming traditional machine learning models in few-shot settings. \cite{Zhu2024Prompting, niu2024ehr} The Health-LLM, a LLM framework that has vision capability in the future integrating health reports and medical knowledge into LLMs, has also been proposed for enhanced disease prediction and personalized health management, showcasing its superior performance over existing systems \cite{Mingyu2024Health}.

Despite the potentials, LLM and MLLMs face limitations in clinical decision support. Explainability and interpretability remain challenging, as their complex decision-making processes often lack transparency, hindering clinician trust and understanding \cite{Park2024Assessing}. Another concern is the potential for bias and unfairness due to inherent biases in MLLM training data, which can exacerbate healthcare disparities \cite{Schmidgall2024AgentClinic}. Extensive real-world validation in diverse clinical settings is crucial to ensure the effectiveness and safety of MLLMs before widespread adoption, addressing potential risks and unexpected outcomes \cite{Ong2024Ethical}.

Several real-world case studies have demonstrated the potential of LLMs in clinical decision support. One study showed that an LLM optimized for diagnostic reasoning improved clinicians' differential diagnosis accuracy on challenging medical cases \cite{Daniel2023Towards}. Another study found that an LLM could accurately classify patient acuity levels in the emergency department, comparable to human physicians \cite{Williams2024Use}.

The performance of LLMs in clinical decision support is often evaluated using traditional metrics like accuracy, precision, recall, F1 score, and AUC \cite{Junda2024JMLR}. Evaluating their effectiveness in this context requires moving beyond accuracy and considering additional factors like interpretability, fairness, impact on clinical workflows, and user trust \cite{Kilian2024Review}. The development of standardized evaluation frameworks and benchmarks, such as CLUE and BenchHealth, is crucial for assessing the clinical performance and real-world applicability of LLMs \cite{Amin2024CLUE, Liu2024Large, Congyun2024RJUA}.

\subsection{Medical Imaging}

MLLMs are rapidly transforming medical imaging by offering potential for significant improvements in diagnosis, treatment planning, and patient care. These models, capable of processing and interpreting both textual and visual data, allow for a more comprehensive understanding of patient conditions. The MISS framework, proposed by Chen et al., treats medical Visual Question Answering (VQA) as a generative task, achieving excellent results with fewer multimodal datasets and demonstrating the advantages of generative models in practical applications \cite{Jiawei2024MISS}. A key strength of MLLMs lies in their ability to analyze medical images in conjunction with textual information such as radiology reports, clinical notes, and patient history. This integration of multimodal data allows for a more holistic and nuanced understanding of a patient's condition. Yildirim et al. demonstrate the value of this approach in radiology, arguing that integrating multimodal data can lead to a more comprehensive patient assessment \cite{Yildirim2024Multimodal}. MLLMs can automate the generation of radiology reports, potentially improving efficiency and accuracy while reducing radiologists' workload \cite{Bhayana2024Chatbots}. MLLMs also facilitate visual question answering, enabling clinicians to interact with medical images by asking specific questions and receiving relevant information from the model \cite{Mehandru2024Evaluating}. 

Despite advantages, several limitations hinder the adoption of MLLMs in medical imaging. One major challenge is the reliance on high-quality, labeled data. Chen et al. address this issue in their work on the MISS framework, proposing solutions for leveraging limited datasets \cite{Jiawei2024MISS}. Interpreting complex medical images may requires specialized knowledge that current MLLMs do not fully possess. Mehandru et al. stress the need for high-fidelity simulations to accurately assess LLM performance in these complex scenarios \cite{Mehandru2024Evaluating}. Ethics about fairness in image interpretation is also crucial, as these models can perpetuate existing healthcare disparities if not carefully designed and evaluated. Yildirim et al. discuss these considerations in detail, focusing on design requirements for ethical AI use in radiology \cite{Yildirim2024Multimodal}.

\subsection{Patient Engagement and Communication}

MLLMs has changed patient engagement and communication in healthcare. By integrating visual and textual modalities, MLLMs can create more personalized and interactive experiences, enhance patient education, facilitate communication, and provide tailored health recommendations.

\textbf{Chatbots and Virtual Assistants}: One of the most promising applications of MLLMs in patient engagement is chatbots and virtual assistants. Traditional chatbots often rely on rule-based systems or simple ML models, with limited ability to understand complex queries and provide nuanced responses. MLLMs, however, can understand both text and images to create more natural and engaging conversations and result in improved patient experiences \cite{mihalache2024interpretation, sharma2024depression, niu2024ehr}.

\textbf{Personalized Health Recommendations}: MLLMs can also be used to generate personalized health recommendations by analyzing patient data and medical knowledge. By integrating information from electronic health records, medical literature, and even patient-provided images, MLLMs provide tailored advice on lifestyle changes, medication adherence, etc \cite{dao2024llm}.

\textbf{Patient Education}: Educating patients improves health outcomes, but traditional methods often rely on static materials that may be difficult to understand or hard to be tailored towards individual needs. MLLMs generates personalized educational materials that are interactive, engaging, and easy to comprehend. The MedSumm framework, for example, utilizes LLMs and VLMs to generate detailed summaries of Hindi-English code-mixed medical queries, integrating visual aids to improve comprehension and support personalized medical care \cite{Akash2024MedSumm}. 

\subsection{Research and Development}
MLLMs are offering promising solutions for literature review, drug discovery, clinical trial matching, and knowledge extraction. They have accelerated discoveries and enhanced knowledge extraction.

\textbf{Literature Review and Knowledge Extraction}: MLLMs are proving invaluable for navigating and synthesizing the vast and ever-growing body of biomedical literature. For instance, BioLORD-2023 integrates LLMs with knowledge graphs to achieve state-of-the-art performance in semantic textual similarity, concept representation, and named entity linking, enabling researchers to extract meaningful insights from complex medical texts \cite{Francois2023BioLORD, niu2024large}. Similarly, MedMT5 tries to overcome language barriers in medical research by offering a robust, open-source, multilingual model for the medical domain, allowing for broader access to knowledge across different languages \cite{Iker2024MedMT5}.

\textbf{Drug Discovery}: While still in the early stages, MLLMs in drug discovery holds potential \cite{steurer2024multimodal}. These models can analyze complex biological data, such as protein structures and drug interactions, to identify potential drug targets and accelerate the drug development process. By integrating information from various modalities, MLLMs can facilitate a more holistic understanding of disease mechanisms and drug interactions, potentially leading to the discovery of novel therapeutics and personalized medicine approaches.

\textbf{Clinical Trial Matching}: MLLMs can significantly improve the efficiency and accuracy of matching patients to suitable clinical trials. These models can analyze patient data, including medical history, genetic information, and imaging data, to identify potential eligibility criteria and match patients with ongoing trials. The ability of MLLMs to process and understand multimodal data can enhance the identification of eligible patients, leading to more effective recruitment and potentially faster clinical trial completion.

\subsection{Administrative Tasks}

Administrative tasks in healthcare is immense, which consumes significant time and resources that could be allocated to improve patient care. MLLMs offer transformative solutions by automating many of these tasks, which streamlines processes and improves the overall efficiency. MLLMs can handle tasks in documentation, billing, scheduling, etc. with remarkable speed and accuracy.

\textbf{Automation of Documentation}: MLLMs are transforming clinical documentation by automating tasks such as generating radiology reports \cite{Nakaura2024impact} and transcribing medical conversations \cite{Ayo2024Sound}. This automation can free up clinicians' time, allowing them to focus on patient care rather than paperwork. For example, one study explored the use of LLMs to simplify radiological reports for improved patient comprehension, finding that LLMs can effectively create more accessible reports while acknowledging the need for careful validation to mitigate potential inaccuracies \cite{Artsi2024Large}.

\textbf{Billing and Scheduling}: The application of MLLMs in billing and scheduling processes can significantly improve efficiency and reduce errors. These models can analyze patient data, insurance information, and scheduling constraints to automate appointment scheduling, generate billing codes, and process insurance claims. By streamlining these tasks, MLLMs can reduce administrative burdens on healthcare staff and improve patient satisfaction by reducing wait times and simplifying billing processes.

\section{Research Gaps and Unanswered Questions}

\subsection{Data Limitations and Needs}

While the potential of MLLMs in healthcare is significant, their development and evaluation are hindered by limitations in data resources. As highlighted by \cite{Tu2023Towards}, medicine is inherently multimodal, with data spanning text, imaging, genomics, and more. Yet, current research faces several key challenges related to data:

\textbf{Scarcity of Large-Scale, Multimodal Datasets}: Existing biomedical datasets are often limited in size and scope, particularly those incorporating multiple modalities. Some researchers mitigates the lack of datasets with locally-aligned phrase grounding annotations for complex semantic modeling \cite{Boecking2022Making}, while other researchers often propose new dataset when releasing new models \cite{Sheng2023BiomedCLIP}. The lack of large-scale datasets and their restricted size and scope is a major bottleneck for training robust and generalizable MLLMs for diverse medical tasks, especially when considering the need for datasets that reflect real-world clinical scenarios \cite{Mingyu2024Health, Mehandru2024Evaluating}.

\textbf{Lack of Diversity and Representation}: Existing datasets often lack diversity in terms of patient demographics, medical conditions, and healthcare settings. This issue is particularly relevant when considering the potential biases introduced \cite{Schmidgall2024Addressing}. Chen et al. emphasizes the challenges of few-shot learning in predicting rare disease areas due to limited data \cite{Zekai2023Language}. The lack of representation results in biased models that perform poorly on underrepresented populations or specific medical conditions. The reliance on single-language data, primarily English, is also a major concern \cite{Jin2024Better, Qiu2024Towards, Iker2024MedMT5}. It is difficult to access large amounts of domain-specific pre-training data for multiple languages, which makes it difficult to resolve linguistic bias \cite{Iker2024MedMT5}.

\textbf{Challenges in Data Acquisition and Annotation}: Obtaining high-quality, annotated multimodal data in healthcare is complex and resource-intensive. \cite{Khare2021MMBERT} notes that medical image annotation is costly and time-consuming, while \cite{Shah2023Creation} points to the lack of LLMs trained on medical records. This challenge is complicated by the need for expert annotations \cite{He2023Survey, Wu2024PMC}. The scarcity of medical image-text pairs for pre-training, due to privacy and cost issues, is another major issue \cite{Jiawei2024MISS}. Additionally, ensuring data privacy and obtaining informed consent are critical ethical considerations that require careful attention, particularly when dealing with sensitive medical information \cite{Ong2024Ethical}.

\subsection{Interdisciplinary Collaboration and Knowledge Integration}

\subsubsection{Fostering Effective Interdisciplinary Collaboration}

The development of clinically relevant and useful MultiModal Large Language Models (MLLMs) requires bridging the gap between computer science and medicine. This interdisciplinary challenge calls for collaboration among medical professionals, data scientists, ethicists, and policymakers \cite{Clusmann2023future, Bean2023Exploring, Park2024Assessing}. Such collaboration is essential to foster a shared understanding of both the technical capabilities of LLMs and the specific needs and constraints of the healthcare domain.

As LLMs become more integrated into healthcare workflows, it is crucial to define the roles and responsibilities of various stakeholders \cite{Francois2023Self}. \cite{Harrer2023Attention} stresses the importance of incentivizing users, developers, providers, and regulators to prepare for the transformative role of LLMs in evidence-based sectors. This preparation includes establishing clear guidelines for accountability and oversight to ensure the safe and ethical use of these powerful tools in healthcare settings.

Clinicians' expertise is vital in guiding the development and evaluation of MLLMs to ensure they address real-world clinical needs\cite{D2023Deciphering, Daniel2023Towards} illustrates how integrating an LLM optimized for diagnostic reasoning into a clinical workflow can improve diagnostic accuracy and comprehensiveness. However, further research is needed to explore effective methods for incorporating clinicians' feedback and domain expertise throughout the model development process. This ongoing collaboration between healthcare professionals and AI developers is key to creating MLLMs that can truly enhance patient care and clinical decision-making \cite{Wenhao2024Multimodal, Malek2023Toward, Armitage2024Large, Briganti2023clinician}.

\subsection{Enhancing Knowledge Integration}

Beyond textual data, integrating domain-specific knowledge from sources like medical ontologies, knowledge graphs, and clinical guidelines is essential for the effectiveness of Large Language Models (LLMs) in complex medical tasks \cite{Boecking2022Making}. A Significant challenge in deploying LLMs for healthcare is addressing the issues of hallucinations and bias. LLMs can generate factually incorrect information and perpetuate biases present in their training data, which is particularly concerning in medical contexts. To tackle this problem, \cite{Logesh2023Med} introduces Med-HALT, a benchmark and dataset specifically designed to evaluate and mitigate hallucinations in medical LLMs. This tool emphasizes the critical need to address these issues for safer healthcare applications. Additionally, \cite{Guangyu2023ClinicalGPT} underscores the importance of incorporating diverse real-world data and domain-specific knowledge to reduce factual inaccuracies and improve the model's grounding in real-world clinical scenarios.

The development of multilingual models represents another crucial area for advancement in medical LLMs. Most LLMs are trained primarily on English data, which limits their accessibility and applicability in diverse linguistic contexts. The potential of bilingual fine-tuned LLMs, such as Taiyi, can achieve superior performance on biomedical NLP tasks compared to general LLMs \cite{Luo2024Taiyi}. However, more research is needed to develop effective methods for creating and evaluating multilingual medical LLMs that can cater to the needs of diverse patient populations \cite{Jin2024Better, Grote2024paradigm}.

\subsection{Ethical and Regulatory Framework}

The potential of MLLMs in healthcare is clear, but their deployment in real-world clinical settings presents significant ethical and regulatory challenges that demand careful consideration and further research.

A key issue is the lack of clear guidelines and regulations specifically tailored for the development, deployment, and evaluation of LLMs in healthcare \cite{Shah2023Creation, Clusmann2023future}. Existing frameworks for medical AI may not fully address the unique ethical and legal implications of LLMs, especially in the context of multimodality. This gap in comprehensive guidance creates uncertainty for developers, clinicians, and regulators, which could impede responsible innovation and safe implementation.

Bias, fairness, and transparency are critical concerns in the use of LLMs in healthcare. Several studies highlight the potential for bias due to imbalances in training data \cite{Harrer2023Attention, Zekai2023Language, Jiang2024754}. This can result in unfair or inaccurate outcomes, particularly for underrepresented or marginalized populations. The lack of transparency in LLM and MLLM training processes and decision-making mechanisms also raises concerns about accountability and trust \cite{Clusmann2023future}. Future research should focus on developing robust methods for identifying, quantifying, and mitigating biases, as well as ensuring transparency in their development and deployment.

Patient privacy and data security are important when using LLMs in healthcare, as they involve processing sensitive patient information. Integrating multiple data modalities in MLLMs adds complexity to data management and raises additional privacy concerns \cite{Acosta2022Multimodal}. Developing secure data storage, de-identification techniques, and access control mechanisms are crucial areas for future research. These challenges are particularly evident in specific medical fields, such as dentistry, where the use of LLMs requires robust safeguards to protect patient data \cite{Wang2023Accelerating, Huang2023ChatGPT, Hu2024Advancing}.

Additional research is needed to define the optimal balance between human oversight and LLM autonomy, and establishing robust governance structures for LLMs in healthcare is essential to ensure accountability and public trust. A framework for evaluating LLMs in healthcare, including a governance layer to ensure accountability and public confidence, has been proposed \cite{Reddy2023Evaluating}. Clear guidelines and standards are needed for data governance, model development, performance evaluation, bias mitigation, and transparency. A collaborative approach involving developers, clinicians, ethicists, regulators, and patients is vital for establishing trust and promoting the responsible use of LLMs in healthcare \cite{Youssef2023Importance}.

\subsection{Technical Advancements Required}

Realizing MLLMs' potential requires overcoming significant technical challenges. Existing research highlights several key areas where advancements are urgently needed.

\textbf{Advancing Modality Alignment Methods}

Current modality alignment methods, which aim to bridge the semantic gap between different data types like text and images, often struggle to capture the complex relationships and nuances present in medical data. This limitation hinders the ability of MLLMs to integrate information effectively and generate accurate and coherent outputs \cite{Jun2024Large, Van2024On}. Novel approaches are needed to create more robust and nuanced alignment methods that can capture the complex interdependencies between different modalities, ensuring a more holistic understanding of medical data.

\textbf{Unveiling the ``Black Box"}

The "black box" nature of large language models, where their internal workings and decision-making processes remain opaque, is a significant challenge for their deployment in high-stakes medical decisions. Clinicians need to understand the rationale behind AI-generated outputs to trust and validate their recommendations. The lack of transparency and interpretability in current LLMs hinders the ability to identify potential biases, errors, or inconsistencies in their reasoning. Further research to understand how LLMs make decisions, particularly in the context of assessing clinical acuity is needed \cite{Williams2024Use}. Developing methods to make LLMs more transparent and interpretable is crucial for ensuring their safe and responsible use in medical applications \cite{Alsentzer2023Zero, Aleksa2023Interpretable}.

\textbf{Enhancing Generalization and Robustness}

Achieving reliable generalization and robustness across diverse medical contexts, patient populations, and languages is crucial for the real-world deployment of MLLMs. Current models often struggle to generalize beyond their training data, which leads to inaccuracies and biases when applied to new populations or scenarios. The study by Zhang et al. demonstrates that while LLMs can effectively analyze data from specific medical specialties, their performance often decreases when applied to other areas \cite{Mehandru2024Evaluating}. Additional efforts should focus on developing methods to enhance the generalization capabilities of MLLMs, ensuring that they perform consistently and reliably across different medical contexts, diverse patient populations, and various languages.

\textbf{Developing Efficient and Scalable Models}

The large size and computational demands of MLLMs pose a significant barrier to their deployment in resource-constrained settings. Training and deploying these models require substantial computational power, which can be prohibitively expensive \cite{Chunyuan2023LLaVA, Bi2024AI}. Developing efficient and scalable models that operates on less powerful devices or with reduced computational resources is crucial for making these technologies more accessible and equitable in healthcare.

\section{Future Directions and Conclusion}
This review has emphasized the potential of LLMs and MLLMs to revolutionize medicine and healthcare. While showing early promise in areas like patient-trial matching \cite{Dyke2024Autonomous}, generating radiology reports \cite{Liu2023Radiology}, and assisting with clinical diagnostics \cite{Tu2023Towards}, LLMs are still in their early stages. Significant research gaps remain and must be addressed to unlock their full potential and ensure safe, responsible, and equitable integration into clinical practice.

\subsection{Data Augmentation and Access}
\textbf{Data Augmentation and Access}: The scarcity of large-scale, high-quality, and diverse multimodal datasets is a major bottleneck \cite{Chunyuan2023LLaVA}. This is especially true for languages other than English \cite{Iker2024MedMT5}. Future research shall focus on:

\textbf{Dataset Creation and Curation}: Developing large, well-annotated datasets encompassing diverse medical specialties, patient populations, and languages is crucial \cite{Qiu2024Towards}. This includes incorporating visual data like medical images, alongside text from EHRs, clinical notes, and medical literature \cite{Xie2024Me}. Datasets should represent real-world scenarios and address issues like imbalanced data \cite{Meng2024application}. The OmniMedVQA benchmark provides a good example of a comprehensive dataset that addresses some of these challenges \cite{Yutao2024OmniMedVQA}.

\textbf{Privacy-Preserving Data Sharing}: Investigating innovative methods like federated learning \cite{Wang2023Accelerating} to enable collaborative data sharing and model training while preserving patient privacy.
    
\textbf{Standardization and Interoperability}: Developing standardized data formats and categories to facilitate data integration and interoperability across different healthcare systems and institutions. This is crucial for training models that can generalize well to new settings \cite{Longwell2024Performance}.

\subsection{Advanced Modality Alignment}
A key area for future research is developing more sophisticated methods for aligning different modalities. Future research could focus on developing novel architectures and training strategies that can better capture the complex relationships between different modalities, leading to more accurate and robust predictions \cite{Meiqi2024Quantifying}. 

\subsection{Interpretability and Explainability}

A critical area for future research is enhancing the interpretability and explainability of MLLMs. This lack of transparency in current LLMs MLLMs can hinder trust and adoption in clinical settings. Traditional evaluation methods for MLLMs are usually insufficient for clinical settings, as they don't adequately assess their impact on real-world workflows \cite{Mehandru2024Evaluating}. Future research should focus on developing methods to make MLLM decision-making processes more transparent and understandable, such as generating human-readable explanations for their predictions or visualizing the processes that contribute to their decisions.

\subsection{Robust Evaluation Frameworks}
Robust and standardized evaluation frameworks becomes increasingly critical as MLLMs become increasingly sophisticated. Current evaluation methods often rely on limited datasets and metrics to non-clinical tasks, restricting the potential to capture the full range of capabilities and biases \cite{Huang2024Comprehensive}. To ensure the safe and effective of clinical MLLMs, future effort should spend on developing additional standardized benchmarks with closer clinical relevance.

\subsection{Ethics and Compliance}

The ethical implications of MLLMs in healthcare cannot be overstated. Its training data contain sensitive patient information, raising concerns about privacy and data security \cite{peng2024securinglargelanguagemodels}. Biased training data can lead to discriminatory outcomes, potentially exacerbating existing health disparities. Therefore, clear regulatory frameworks and guidelines are necessary to govern the development, deployment, and use of MLLMs in clinical settings \cite{Ong2024Ethical}. Addressing these ethical and compliance challenges will be beneficial to establish trust and ensure the responsible use of MLLMs in healthcare.

\bibliographystyle{IEEEtran}  
\bibliography{references} 


\end{document}